# Disorder-induced time effect in the antiferromagnetic domain state of $Fe_{1+y}Te$

Jan Fikáček[1*], Jonas Warmuth[2], Fabian Arnold[3], Cinthia Piamonteze[4], Zhiqiang Mao[5], Václav Holý[6], Philip Hofmann[3], Martin Bremholm[7], Jens Wiebe[2], Roland Wiesendanger[2], and Jan Honolka[1]

[1]*Center for Analysis of Functional Materials, Institute of Physics of CAS, Na Slovance 1999/2, 182 21 Prague 8, Czech Republic*
[2]*Department of Physics, Hamburg University, Jungiusstraße 9A, D-20355 Hamburg, Germany*
[3]*Department of Physics and Astronomy, Interdisciplinary Nanoscience Center (iNANO), Aarhus University, Ny Munkegade 120, 8000 Aarhus C, Denmark*
[4]*Paul Scherrer Institut, Forschungsstrasse 111, 5232 Villigen PSI, Switzerland*
[5]*Department of Physics, Pennsylvania State University, University Park, PA 16802, USA*
[6]*Charles University, Faculty of Mathematics and Physics, Department of Condensed Matter Physics, Ke Karlovu 5, 12116 Prague 2, Czech Republic*
[7]*Department of Chemistry, Interdisciplinary Nanoscience Center (iNANO), Aarhus University, Langelandsgade 140 8000 Aarhus C, Denmark*

**Abstract**

We report on temperature-dependent soft X-ray absorption spectroscopy (XAS) measurements utilizing linearly polarized synchrotron radiation to probe magnetic phase transitions in iron-rich $Fe_{1+y}Te$. X-ray magnetic linear dichroism (XMLD) signals, which sense magnetic ordering processes at surfaces, start to increase monotonically below the Néel temperature $T_N$ = 57 K. This increase is due to a progressive bicollinear antiferromagnetic (AFM) alignment of Fe spins of the monoclinic $Fe_{1+y}Te$ parent phase. This AFM alignment was achieved by a [100]-oriented biasing field favoring a single-domain state during cooling across $T_N$. Our specific heat and magnetization measurements confirm the bulk character of this AFM phase transition. On longer time scales, however, we observe that the field-biased AFM state is highly unstable even at the lowest temperature of $T$ = 3 K. After switching off the biasing field, the XMLD signal decays exponentially with a time constant $\tau$ = 1506 s. The initial XMLD signal is restored only upon repeating a cycle consisting of heating and field-cooling through $T_N$. We explain the time effect by a gradual formation of a multi-domain state with 90º rotated AFM domains, promoted by structural disorder, facilitating the motion of twin-domains. Significant disorder in our $Fe_{1+y}Te$ sample is evident from our X-ray diffraction and specific heat data. The stability of magnetic phases in Fe-chalcogenides is an important material property, since the $Fe(Te_{1-x}Se_x)$ phase diagram shows magnetism intimately connected with superconductivity.

*Email address of the corresponding author: janfikacek@gmail.com

## 1. Introduction

Iron-based chalcogenides (Chs) are van der Waals crystals consisting of stacked quasi-two-dimensional FeCh layers. Their crystal structure is the simplest one among iron-based superconductors making them ideal candidates for both experimental and theoretical investigations. In this group of materials, FeTe and FeSe, and their respective substitution system Fe(Te$_{1-x}$Se$_x$), have turned out to be very useful for studies of the interplay between magnetism and superconductivity. FeSe is superconducting below a critical temperature $T_c$ = 8 K [1], which is tunable by various parameters and can be increased up to 100 K when a monolayer of FeSe is grown on SrTiO$_3$ [2]. On the contrary, the analog tellurium-based compound is non-superconducting. Depending on the amount of excess iron atoms $y$, Fe$_{1+y}$Te has different antiferromagnetic (AFM) ground states and crystal structures [3-7]. For $y$ < 0.11, the room temperature tetragonal crystal structure (P4/*nmm*) is replaced by a monoclinic one upon cooling when Fe$_{1+y}$Te orders antiferromagnetically below $T_N$ = 57 K [3-6]. For higher amounts of excess iron (0.11 < $y$ < 0.13), the transitions are split in temperature. First, a complex magnetic phase sets in together with a transition to an orthorhombic crystal structure (*Pmmn*). At lower temperature, it is followed by a broad transition leading to a mixture of orthorhombic and monoclinic crystal structures down to the lowest temperatures state accompanied by a spin-density wave antiferromagnetic ordering [5, 8]. Finally, for $y$ > 0.13, there appears again only one transition to an orthorhombic incommensurate AFM phase. Only in the monolayer limit, FeTe also shows superconducting properties below $T$ = 6 K when placed on top of Bi$_2$Te$_3$ [9, 10]. Recently, 100 nm wide and several 100 nm long AFM twin domains have been reported to form on the surface of bulk FeTe as a result of a strong magneto-elastic coupling [11]. Explaining the large difference between FeSe and FeTe ground states (superconducting vs. antiferromagnetic) despite their almost alike crystal structures, might help to understand the origin of unconventional superconductivity.

Twinning processes under the influence of strong magneto-elastic coupling usually lead to complicated time dependent magnetic properties, e.g. observed in shape memory alloys. Also iron-based superconductors have been reported to show time effects in their magnetic properties. Generally, time dependent phenomena of magnetic properties are usually observed in systems where the thermodynamic equilibrium is not fully reached. For example in spin glass states, this is caused by the fact that after zero field cooling, the realized magnetic state does not have the lowest energy [12]. As a result, the system attempts to find a state with lower energy. Typically, this is connected with a decay of the bulk magnetization on a time scale of hours [12, 13]. Interestingly, FeTe has a spin glass ground state when Te is substituted by Se [14] or Fe in Fe$_{1.1}$Te by Cu [15]. However, for pure Fe$_{1+y}$Te, a spin-glass-like behavior has not been observed so far.

In this work, we report on an analogous behavior to spin glasses when probing surface magnetic properties of Fe$_{1+y}$Te with a nominal excess iron concentration $y$ = 0.073. Magnetism was studied via soft X-ray absorption spectroscopy (XAS) utilizing linearly polarized synchrotron radiation tuned to the energies around the Fe $L_3$ and $L_2$ edges. Due to the presence of a strong magneto-crystalline coupling between structural and magnetic properties on a microscopic scale, the magnetic ordering is accompanied by a formation of at least two different magneto-structural domains [16] reminiscent of a classic domain formation observed in ferromagnetic materials. Analogically, we will show that the domain formation can be suppressed by cooling across $T_N$ in the presence of a sufficiently high magnetic field applied along an *a*-axis of the room temperature tetragonal structure. Magnetic domains then have the same axis orientation down to the lowest temperature of $T$ = 3 K reached in our experiment. However, despite the low temperatures, the multi-domain state is gradually restored after a few tenths of minutes as seen by an exponential decay of the X-ray magnetic linear dichroism signal (XMLD). After heating up above $T_N$ and cooling back with the same magnetic field, the XMLD signal is fully recovered proving that the process is reproducible.

Based on a thorough structural, magnetic, and specific heat analysis of the samples, we attribute the restoration of a twinned state at low temperatures of a few Kelvin to the presence of structural imperfections, which can lower energy barriers for a multi-domain state formation in $Fe_{1+y}Te$.

## 2. Results and discussion

$Fe_{1+y}Te$ single crystals were synthesized using the flux method where the excess iron concentration $y$ was kept as low as possible [17]. Before measuring XAS, we characterized the structural properties of our samples by measuring single crystal X-ray diffraction. The pattern is shown in Fig. 1 (a). The data is compared with a fit based on reported data for $Fe_{1.087}Te$ with $a = 3.821(9)$ Å, $c = 6.285(5)$ Å [18]. The peaks are much wider than what is expected from the fit, pointing to an increased disorder on the local scale [see the zoomed peak in Fig. 1 (a)]. Except for the peaks attributed to the $Fe_{1.087}Te$ crystal structure, we found additional peaks (marked by two vertical arrows) from the silver paste used for sticking the single crystal to its holder. In addition, a comparable single crystal from the same batch was also probed via specific heat by a relaxation time method as adopted by PPMS instruments (Quantum Design) and by magnetization measurements using a SQUID magnetometer. We observed two peaks related to two consecutive transitions at $T = 54$ K and 57 K when heating the sample [see Fig.1 (b)] similar to those reported in Ref. [5] for $Fe_{1.115}Te$. The peak at $T = 54$ K coincides with the onset of the AFM transition seen as a step in our bulk magnetization data ($M/H$) [Fig. 1(c)] when approaching the Néel temperature from below. For cooling, the second (1st order) transition is more smeared out, thus not observable in our specific heat data. Considering the fact that we also observed a hysteresis in $M/H$, we conclude that the majority phase of our sample has actually a slightly higher excess iron concentration between 0.115 and 0.12 [5]. In the specific heat data, an additional minor peak appears at around $T = 69.5$ K, which is most probably related to a different $Fe_{1+y}Te$ ($y < 0.06$) phase present as an impurity and having a different stoichiometry from the majority phase. Compared with transition temperatures in the previously published phase diagram [5, 6], this impurity phase should contain a lower amount of excess iron, most likely $y < 0.06$. This is supported by the fact that at the very same temperature, magnetization starts to decrease as a result of the impurity phase AFM ordering. Another anomaly appears at around $T = 47$ K possibly due to a domain reorientation in the vicinity of the Néel temperature.

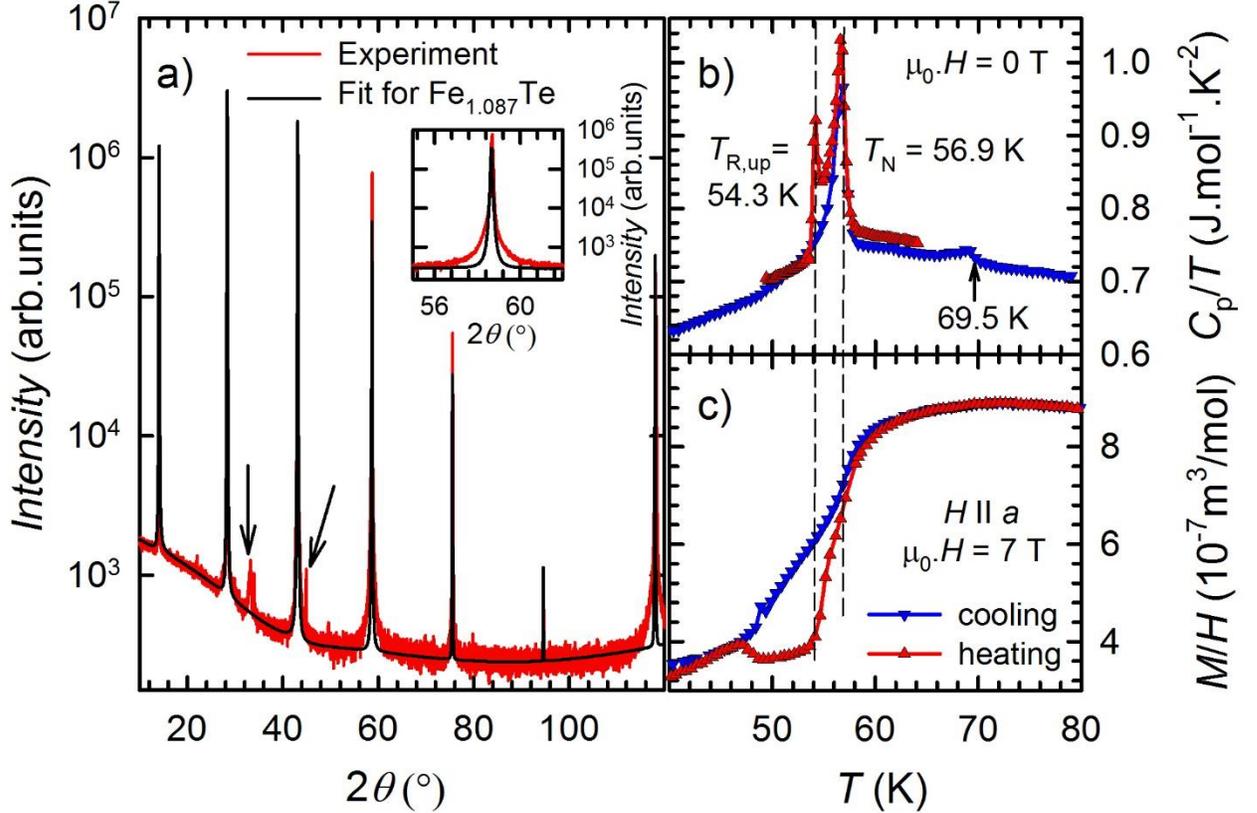

Fig. 1: (a) X-ray diffraction pattern of $Fe_{1.073}Te$ on a logarithmic scale fitted by the structural parameters reported for $Fe_{1.087}Te$ [18]. Additional peaks coming from silver paste are marked by two arrows. The inset shows a magnified region around $2\theta = 58°$. (b) Specific heat data measured at 0 T field during cooling and heating. Two vertical dashed lines mark phase transitions related to the majority phase. The arrow at $T = 69.5$ K points at an anomaly caused by a minor second phase, $Fe_{1+y}Te$. (c) Rescaled magnetization ($M/H$) versus temperature measured along the $a$-axis in the field of 7 T.

XAS measurements were performed at the EPFL-PSI *X-treme* beamline at Swiss Light Source in Villigen [19]. The sample pre-characterized by XRD was mounted onto a molybdenum holder with its $c$-axis perpendicular to the holder plane. The azimuthal orientation was verified by low-energy electron diffraction in advance. Since FeTe consists of stacked covalently bonded Te-Fe-Te triple layers where each triple layer is bonded to another by weak Van der Waals forces only, it was possible to in-situ cleave the sample by a scotch tape in order to get a clean surface. Cleaving was done after pumping for a few hours to reach UHV conditions ($10^{-8}$-$10^{-9}$ mbar). After cleaving, the sample was transferred to the XAS measurement chamber immediately.

Photon energy dependent XAS data was measured using the surface sensitive total electron yield method (TEY). First, we checked a selected surface area for oxygen contamination. If the oxygen signal was negligible, we proceeded with measuring XAS around Fe-$L_3$ and -$L_2$ in the energy range 690 eV – 745 eV for linearly polarized light with horizontal ($\varepsilon_\parallel$) and vertical ($\varepsilon_\perp$) directions. We measured the spectra at several temperatures during cooling from room temperature down to the base temperature of 3 K. During each cooling step, a biasing field of 6.8 T was applied as shown in the inset of Fig. 2. XMLD signals defined as $[XAS(\varepsilon_\parallel) - XAS(\varepsilon_\perp)]$ were measured at 0.1T field at each target temperature. For our measuring geometry depicted in the inset of Fig. 2, the XMLD scales as $\sim \cos^2(\alpha)<M^2>$ where $\alpha$ is the angle between the polarization vector $\varepsilon_\parallel$ and the direction of AFM moments projected onto the surface plane [20]. According to that, the maximum XMLD intensity is expected for an AFM state whose surface projection is aligned along [100], and similar to that as reported at low excess Fe < 0.12 [4]. In our experiment, the XMLD signals started to appear at $T$ = 55 K, the first spectrum below $T_N$. Upon further cooling down to the base temperature of 3 K, the magnitude of the XMLD signals was increasing monotonically.

Fig. 2 shows a typical XAS spectrum at $T$ = 3 K averaged out over both polarizations with a peak height defining max-$L_3$. The XAS line-shape is broad and shows a kink at the high energy side of the $L_3$ edge at ~ 709 eV. The absence of sharp multiplet features confirms the metallic character of Fe without signs of oxidation. The respective XMLD signal is also shown in Fig. 2 but magnified by a factor of 5 for better visibility. The finite XMLD signal clearly indicates that a preferential bicollinear AFM state with a strong projection along [100] was achieved by the [100]-oriented biasing field of 6.8 T, favoring a single-domain state during cooling across $T_N$. According to previous studies, much higher magnetic fields of at least 14 T were necessary for a successful detwinning of $Fe_{1.1}$Te in the $a$-$b$ plane at $T$ = 55 K (5 K below $T_N$) [21, 22]. However, we expect that critical fields for an efficient biasing procedure are indeed lower when increasing the temperature close to $T_N$. This is suggested also by recent resistivity measurements on FeTe where field cooling across $T_N$ at a magnetic field of only 2 T resulted in an almost fully detwinned state [23].

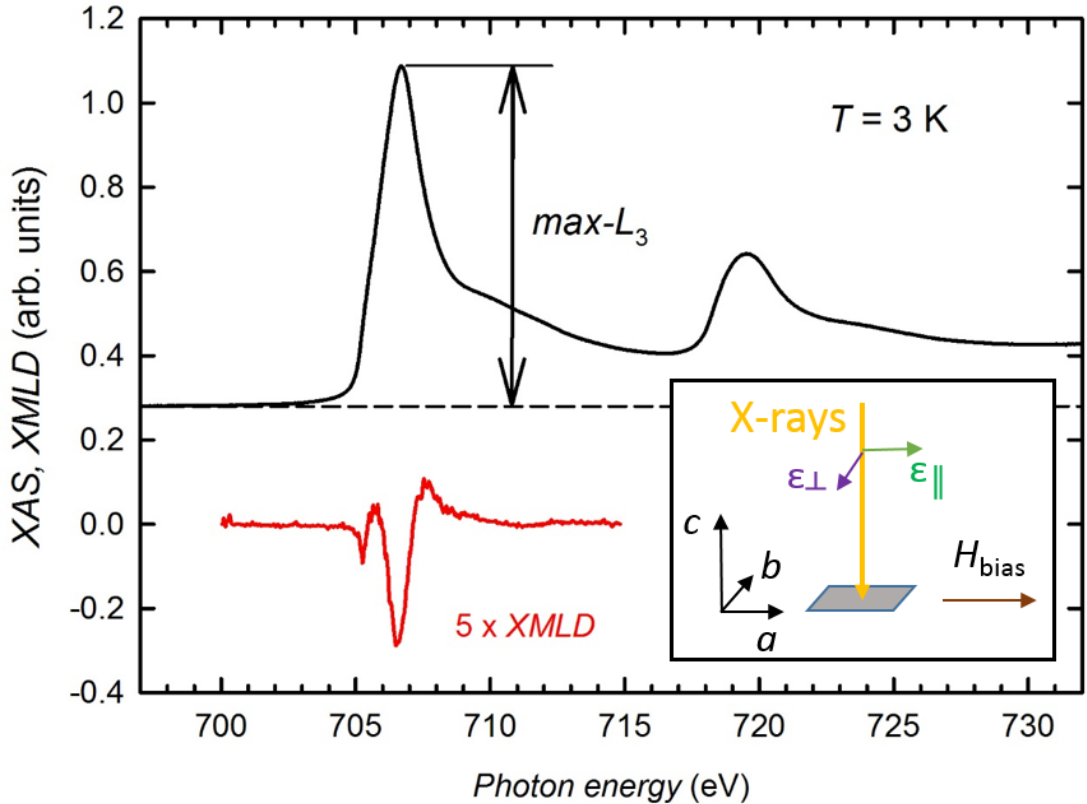

Fig. 2: The top part shows a typical XAS spectrum averaged over horizontal ($\varepsilon_\parallel$) and vertical ($\varepsilon_\perp$) polarizations taken at $T$ = 3 K after cooling in a biasing field of 6.8 T. The inset sketches the experimental geometry. Bottom part: The respective XMLD signal from a restricted energy range after warming up above

$T = 100$ K and cooling back to $T = 3$ K. The XMLD spectrum was magnified by a factor of 5 for better visibility.

After the successful field-biasing procedure, which generated a partially detwinned state at low temperatures, we investigated its stability. On larger time scales, we observed that the magnitude of the XMLD signal was getting smaller. In order to study the kinetics of this effect in more detail, we heated up the sample above $T = 100$ K and repeated the biasing procedure via cooling to $T = 3$ K without interruption. Afterwards, we switched off the biasing field and started collecting polarization dependent XAS data around the $L_3$ edge (700-715 eV) for 90 minutes at constant temperature of $T = 3$ K. The resulting time-dependent XMLD spectra are plotted in Fig. 3. The strongest change of the XMLD spectrum is visible within the first 40 min. At the same time, the minima and the left maxima both gradually shift to lower energy. After 80 minutes, the XMLD spectrum is similar to that measured at high temperatures $T = 70$ K, well above the $T_N$.

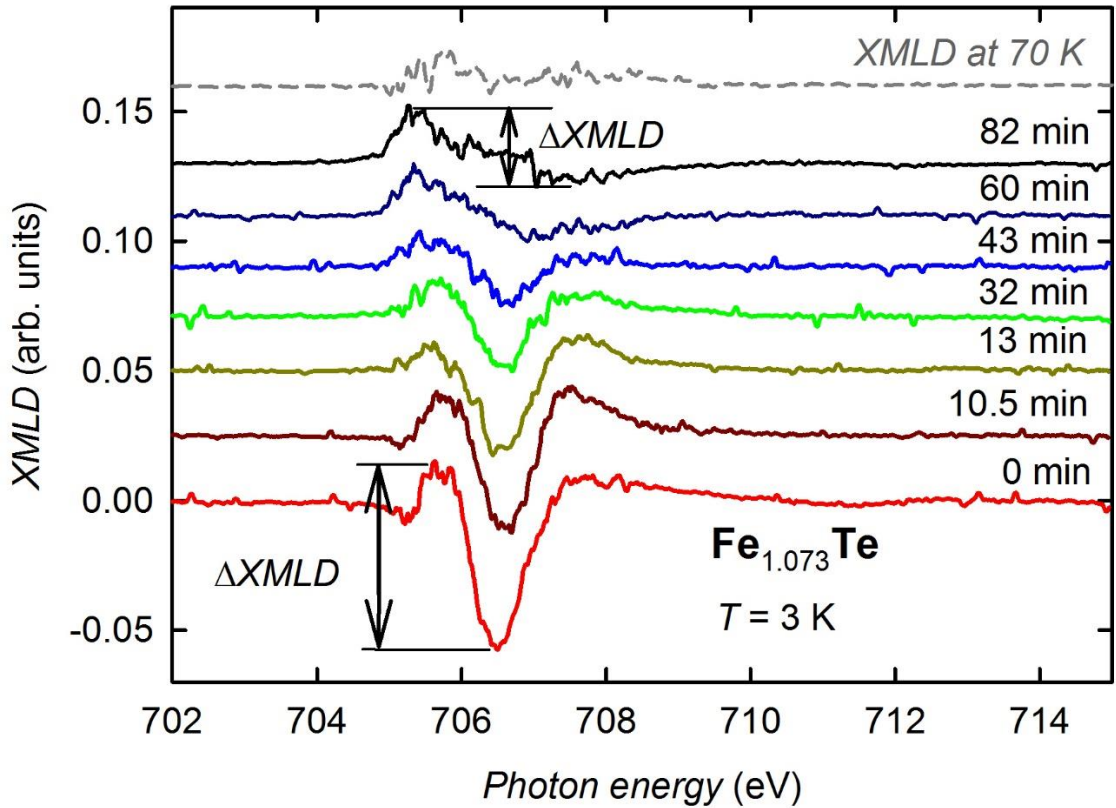

Fig. 3: Decay of the XMLD intensity with time. The spectra were measured after switching off the biasing field while staying at the same sample spot and temperature of $T = 3$ K. For better visibility, only a limited number of spectra displaced by vertical offsets is shown. For comparison, an XMLD signal at $T = 70$ K ($T > T_N$) is shown (dashed line), which was measured during initial cooling.

The magnitude of the dichroism signal, $\Delta XMLD$, was quantified by the difference between the maximum and the minimum of the given XMLD signal located at approximately 705.7 eV and 706.6 eV, respectively, as shown in Fig. 3. An exponential decay of the dichroic signal is clearly observable from Fig. 4, where we plotted the dependence of $\Delta XMLD$ versus time. $\Delta XMLD$ was normalized by the peak height *max-L$_3$* of the Fe-*L$_3$* XAS signal. Since Δ*XMLD* were only in the order of a few percent, error bars are large and vary with the quality of the respective XAS spectra. The decay itself is well described by a relaxation time $\tau = (1506 \pm 200)$ s when assuming an exponential decay shown in Fig. 4. Such thermally reactivated twinning was observed before in magnetoresistance measurements in the related tetragonal (*I*4/*mmm*) Fe-pnictide compound Ba(Fe$_{1-x}$Co$_x$)$_2$As$_2$ with a comparable time constant of 3300 s at $T = 5$ K [13].

In the following, we further discuss the time effect at low temperatures. The tendency of Fe-rich FeTe to form azimuthal domains/twins in the absence of biasing fields was observed in STM measurements on samples with comparable amounts of excess Fe, 8% and 12% [11]. In the chosen experimental XAS geometry, 90º azimuthally rotated AFM domains with a twin boundary running along [110] direction [16] separating domains with interchanged *a*- and *b*-axes would contribute to the XMLD signal with an opposite sign. Thus, they would indeed cancel each other out and compensate the total XMLD signal when twin domains contribute equally within the 30 x 230 μm² (vertical x horizontal) sized XAS probing area defined by the X-ray beam spot. Since in STM measurements on unbiased FeTe samples, domain sizes ranging from at least tens of nm [11] up to hundreds of nm [16] were reported with statistical domain orientations, such a scenario is conceivable.

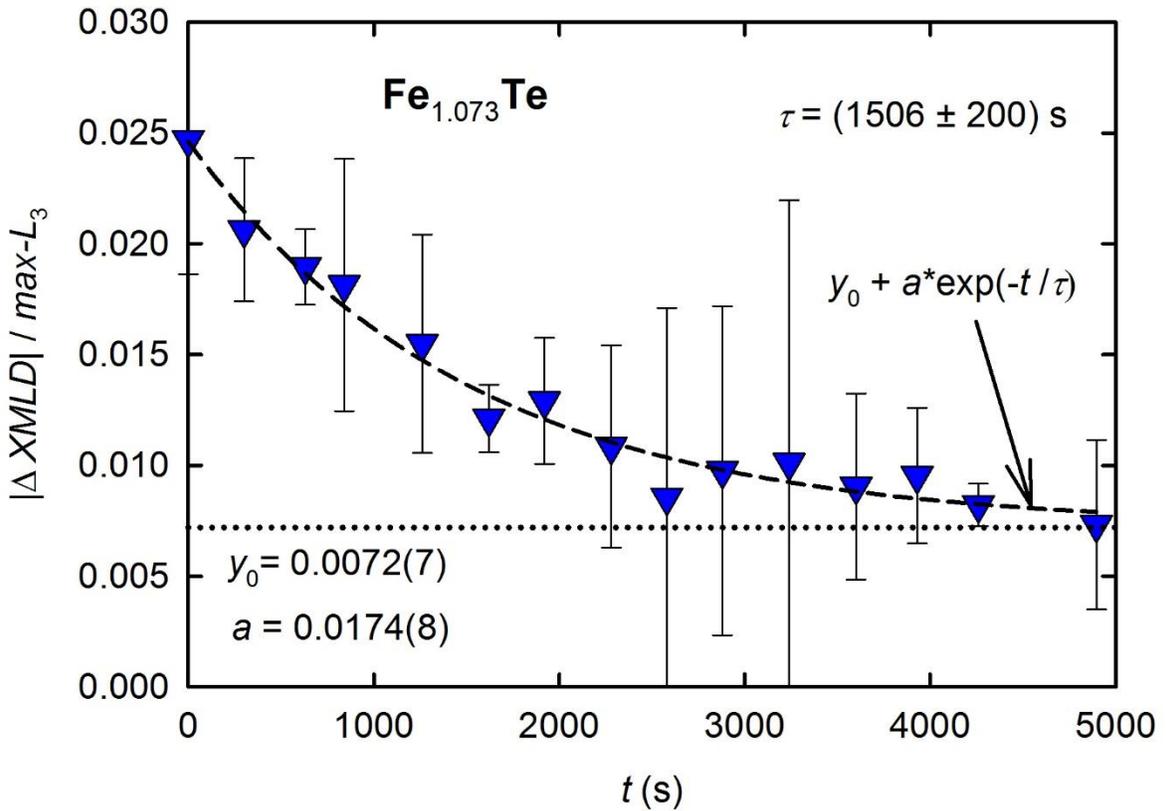

Fig. 4: Time dependence of the magnitude Δ*XMLD* extracted from the experimental XMLD signals in Fig. 3. Δ*XMLD* is shown normalized to *max-L$_3$*. The dashed line corresponds to a fit assuming an exponential decay with a time constant $\tau = (1506 \pm 200)$ s. Details on the fit function and the respective fitted parameters are given in the figure.

The mechanism of domain formation and motion in antiferromagnetic materials, especially those with large magneto-elastic coupling, is less intuitive [24] compared to ferromagnetic materials where magnetic stray field energies generate domain formation processes. One of the early observations of AFM domains in a chalcogenide-containing compound was reported for EuSe a long time ago [25]. Their presence was explained by magnetostriction. Analogically, we can compare it to magnetic shape memory alloys, where these effects also arise due to symmetry breakings having both structural and magnetic origin and leading to a twin formation. In the classic metallic shape memory alloy NiMnGa in a ferromagnetic state, the temporal evolution of twin reorientation via a magnetic field is described by thermally activated creep motion of twin boundaries over a distribution of energy barriers [26]. Hereby, thermally activated magnetoelastic twin boundary motion occurs over a distribution of energy barriers under low-to-moderate driving forces consistent e.g. with a creep model involving short-range local disorder.

In AFM Fe-pnictides with superconducting properties, time effects seem negligible and twin-domain motions are slowed down. In $EuFe_2As_2$, there are two types of domains and the crystal gets detwinned under a relatively low field of 0.1 T applied parallel to the $b$-axis thanks to a strong magnetoelastic coupling [27]. In underdoped $Ba(Fe_{1-x}Co_x)_2As_2$, a residual population imbalance is stable for hours, and possibly longer [28]. Faster time-effects were reported after a partial detwinning in a magnetic field [13].

In the case of the Fe-chalcogenide FeTe, in order to reduce a macroscopic change in the planar dimensions, e.g. at the interface with a sample holder, a multi-domain structure forms to cancel out the difference between the $a$- and the $b$-axis ($\approx 1\%$) [5]. As a result, the single domain state enforced by a field-biasing procedure and under standard experimental conditions (sample holder, glues, etc.) will never possess the lowest energy. Our XMLD measurements prove that the multi-domain state, which is favored from the point of view of elastic strain energies, can be restored after the system is provided a sufficient time to approach its thermodynamic equilibrium. We attribute the instability of the single-domain state to a large structural disorder in our $Fe_{1+x}Te$ sample, which is evident from (i) significantly broadened peaks in its powder X-ray diffraction pattern, and (ii) one extra peak in specific heat located at higher temperature than those expected for the majority phase at $y \approx 0.12$, which we ascribe to a coexisting phase with less excess iron $y \leq 0.06$.

Further evidence of disorder is given by a small remanent magnetization $\approx 2.10^{-3}$ $\mu_B$/per f. u. ($Fe_{1.073}Te$) observed in $M(H)$ scans and persisting up to room temperature (not shown). We assign this ferromagnetic signal to minor ferromagnetic $FeTe_2$ nanoplatelets known to form during sample growth. No anomalies appear in our specific heat measurements at reported phase transition temperatures of large-scale single crystalline $FeTe_2$ [29, 30]. Disorder generally introduces nucleation centers for domain walls, allowing them to restore the multi-domain state on faster time scales, fast enough to be detectable via synchrotron-based XAS experiments under UHV conditions.

Our interpretation regarding the critical importance of disorder for the stability of a detwinned state is underlined by our measurements on a reference sample having an excess iron content of $y = 0.094$. Although XMLD spectra of this sample and of the freshly biased sample with diminishing XMLD are analogous, they are not time dependent. The reference sample was supposedly of higher quality with no traces of an $Fe_{1+y}Te$-like impurity phase, since we did not observe the additional peak at higher temperatures in its specific heat.

Also, the remanent signal in $M(H)$ was found to be $10^{-4}$ μ$_B$/f.u. implying that here, the concentration of FeTe$_2$ impurities is about an order of magnitude smaller than in our Fe$_{1.073}$Te sample.

We want to note that from our XMLD data, we cannot judge if a field-biased AFM state is achieved in the low temperature orthorhombic or monoclinic structure, or their mixture, all appearing close to $y = 0.12$ estimated for the main phase in our sample. For the monoclinic structure, it would not be possible to distinguish between the type of domains having their monoclinic crystal structure rotated around the $a$-axis by $2(\beta - 90°)$ [11] where $\beta$ is the monoclinic angle ($\approx 90.7°$). These domains can be spatially reoriented around the [001] direction above a critical field denoted as $H_{m1}$ in Ref. [16]. For this case, the change in the AFM structure alignment is not as significant as it is for interchanging the $a$- and the $b$-axis [11]. Interestingly, for Fe$_{1.12}$Te it has been found that at high magnetic fields, when a field-polarized paramagnetic phase is induced above $H_{m2}$ [16], a time-dependent retwinning starts when the field is removed. This would be analogical to switching off the magnetic field after finishing our field biasing procedure.

## 3. Conclusions

In summary, we have detected a time dependent decay of a detwinned antiferromagnetic state in in Fe$_{1+y}$Te at temperatures of a few Kelvin by means of X-ray absorption spectroscopy. Based on a thorough structural, magnetic, and specific heat analysis of the samples, we attribute the effect to structural imperfections influencing the stability of magnetic phases in Fe$_{1+y}$Te, especially in states with low entropy and enforced by biasing external magnetic fields thanks to a strong magnetoelastic coupling. These metastable configurations can relax into an energetically favorable twinned state even at low temperatures when enough faults are present. Since the Fe(Te$_{1-x}$Se$_x$) phase diagrams show that magnetism is intimately connected with superconductivity, our results on the dynamics of magneto-structural instabilities of FeTe suggest to extend studies on the interplay of time and disorder also to superconducting FeSe crystals.

**Acknowledgements**

The authors express their highest gratitude to Dr. Sahana Rößler for very helpful discussions. This work was supported by the Czech Science Foundation Grant No. 19-13659S. We acknowledge support by the European CALIPSOplus program for experiments at the Swiss Light Source. This work was supported by the DFG via the priority programme SPP1666 (grant nos. HO 5150/1-2 and WI 3097/2-2). Martin Bermholm and Philip Hofmann acknowledges the Villum Centre of Excellence for Dirac Materials (Grant No.11744) and Martin Bremholm furthermore the Danish Council for Independent Research under the Sapere Aude program (Grant No. 7027-00077B). Zhiqiang Mao acknowledges the support from the US National Science Foundation under grant DMR1707502.